\documentstyle[12pt]{article}

\catcode`\@=11
\long\def\@makefntext#1{
\protect\noindent \hbox to 3.2pt {\hskip-.9pt
$^{{\ninerm\@thefnmark}}$\hfil}#1\hfill}		

 \def\@makefnmark{\hbox to 0pt{$^{\@thefnmark}$\hss}}  
	
\def\ps@myheadings{\let\@mkboth\@gobbletwo
\def\@oddhead{\hbox{}
\rightmark\hfil\ninerm\thepage}
\def\@oddfoot{}\def\@evenhead{\ninerm\thepage\hfil
\leftmark\hbox{}}\def\@evenfoot{}
\def\sectionmark##1{}\def\subsectionmark##1{}}
 
\newcounter{sectionc}\newcounter{subsectionc}\newcounter{subsubsectionc}
\renewcommand{\section}[1] {\vspace{0.6cm}\addtocounter{sectionc}{1}
\setcounter{subsectionc}{0}\setcounter{subsubsectionc}{0}\noindent 
	{\bf\thesectionc. #1}\par\vspace{0.4cm}}
\renewcommand{\subsection}[1] {\vspace{0.6cm}\addtocounter{subsectionc}{1} 
	\setcounter{subsubsectionc}{0}\noindent 
	{\it\thesectionc.\thesubsectionc. #1}\par\vspace{0.4cm}}
\renewcommand{\subsubsection}[1] {\vspace{0.6cm}\addtocounter{subsubsectionc}{1}
	\noindent {\rm\thesectionc.\thesubsectionc.\thesubsubsectionc. 
	#1}\par\vspace{0.4cm}}

\newcounter{appendixc}
\newcounter{subappendixc}[appendixc]
\newcounter{subsubappendixc}[subappendixc]

\renewcommand{\appendix}[1] {\vspace{0.6cm}

        \refstepcounter{appendixc}
        \setcounter{figure}{0}
        \setcounter{table}{0}
        \setcounter{equation}{0}
        \renewcommand{\thefigure}{\Alph{appendixc}.\arabic{figure}}
        \renewcommand{\thetable}{\Alph{appendixc}.\arabic{table}}
        \renewcommand{\theappendixc}{\Alph{appendixc}}
        \renewcommand{\theequation}{\Alph{appendixc}.\arabic{equation}}
        \noindent{\bf Appendix \theappendixc #1}\par\vspace{0.4cm}}

\def\abstracts#1{{
	\centering{\begin{minipage}{30pc}\tenrm\baselineskip=12pt\noindent
	\centerline{\tenrm ABSTRACT}\vspace{0.3cm}
	\parindent=0pt #1
	\end{minipage}}\par}} 
 
\renewenvironment{thebibliography}[1]
	{\begin{list}{\arabic{enumi}.}
	{\usecounter{enumi}\setlength{\parsep}{0pt}
\setlength{\leftmargin 1.25cm}{\rightmargin 0pt}
	 \setlength{\itemsep}{0pt} \settowidth
	{\labelwidth}{#1.}\sloppy}}{\end{list}}
 
\newcounter{itemlistc}
\newcounter{romanlistc}
\newcounter{alphlistc}
\newcounter{arabiclistc}

\newcommand{\fcaption}[1]{
        \refstepcounter{figure}
        \setbox\@tempboxa = \hbox{\tenrm Fig.~\thefigure. #1}
        \ifdim \wd\@tempboxa > 6in
           {\begin{center}
        \parbox{6in}{\tenrm\baselineskip=12pt Fig.~\thefigure. #1}
            \end{center}}
        \else
             {\begin{center}
             {\tenrm Fig.~\thefigure. #1}
              \end{center}}
        \fi}
 
\newcommand{\tcaption}[1]{
        \refstepcounter{table}
        \setbox\@tempboxa = \hbox{\tenrm Table~\thetable. #1}
        \ifdim \wd\@tempboxa > 6in
           {\begin{center}
        \parbox{6in}{\tenrm\baselineskip=12pt Table~\thetable. #1}
            \end{center}}
        \else
             {\begin{center}
             {\tenrm Table~\thetable. #1}
              \end{center}}
        \fi}
 
\def\@citex[#1]#2{\if@filesw\immediate\write\@auxout
	{\string\citation{#2}}\fi
\def\@citea{}\@cite{\@for\@citeb:=#2\do
	{\@citea\def\@citea{,}\@ifundefined
	{b@\@citeb}{{\bf ?}\@warning
	{Citation `\@citeb' on page \thepage \space undefined}}
	{\csname b@\@citeb\endcsname}}}{#1}}
 
\newif\if@cghi
\def\cite{\@cghitrue\@ifnextchar [{\@tempswatrue
	\@citex}{\@tempswafalse\@citex[]}}
\def\citelow{\@cghifalse\@ifnextchar [{\@tempswatrue
	\@citex}{\@tempswafalse\@citex[]}}
\def\@cite#1#2{{$\null^{#1}$\if@tempswa\typeout
	{IJCGA warning: optional citation argument 
	ignored: `#2'} \fi}}

\def\fnt#1#2{\footnotetext{\kern-.3em
	{$^{\mbox{\sevenrm #1}}$}{#2}}}
 
 1
 1
 1

\font\tenbf=cmbx10
\font\tenrm=cmr10
\font\tenit=cmti10

\font\ninerm=cmr9

\begin{document}

\centerline{\tenbf LET THE SPIN AND THE CHARGES  UNIFY }



\baselineskip=16pt
\vspace{0.8cm}
\centerline{\tenrm NORMA MANKO\v C BOR\v STNIK \footnote[1]{Talk
presented at XIX Triangular Meeting on Recent Development in Quantum
Theories, Rome, March 1996 and at IWCQIS 96, Dubna, July 1996.}  }
\vspace{0.3cm}
\baselineskip=13pt
\centerline{\tenit Department of Physics, University of
Ljubljana, Jadranska 19 }
\vspace{0.1cm}
\centerline{\tenit J. Stefan Institute, Jamova 39
Ljubljana , 1001, Slovenia }
\vspace{0.9cm}

\abstracts{
In  space of d ordinary and d Grassmann coordinates, with
$d\ge 15, $ the charges unify with the spin: the Lorentz group $
SO(1,d-1) $ in Grassmann space manifests under certain
conditions as $ SO(1,3) $ (in $ d=4 $ subspace) times $ SO(10)
\supset SU(3) \times SU(2) \times U(1) $ (in the
rest of the space), accordingly the
symmetry group of the S - matrix, which is approximately
unitary in $ d=4 $ ordinary subspace, manifests as the direct
product of the Poincar\' e group in $ d=4 $ subspace and the
groups describing charges.}

\vspace{8mm}



The fact that  ordinary space-time is not sufficient to
describe  dynamics of our world was first recognized in 1925,  
when in addition 
to the infinite dimensional vector space, spanned over 
ordinary coordinate space, 
a space of two vectors - the internal space of the
fermionic spin, was introduced\cite{uhl}.  Since then the
internal space of fermions was enlarged
to be able to describe particle - antiparticle degrees of
freedom 
\cite{dirac} or their handedness (in 1928), the weak
charge\cite{heis} 
 (1932) and the colour charge\cite{gel}
(1964).  The unification of electromagnetic and weak
interactions makes clear that also the
electromagnetic charge origins in the internal 
space\cite{wein}. The internal space of bosons grew more or less
parallel to the internal space of fermions.

Theories connect all symmetries or the coresponding properties,
appearing in physics, with 
appropriate groups and define accordingly quantum numbers: The
spin is connected with the Lorentz group $ SO(1,3). $ Charges
are connected with the group   $U(1)$ ( the electromagnetic symmetry),  
$ SU(2) $ ( the weak symmetry) and   $ SU(3) $ ( the
colour symmetry).

In {\it ordinary} space time  only {\it
vectorial} types 
of representations for the Lorentz group are possible: the
generators of the infinitesimal transformations $ L^{mn} = (x^m
p^n - x^n p^m) $ define intiger angular momenta. 
For the {\it internal} spaces two types of representations for
either the Lorentz group or the groups describing charges are
required: the {\it fundamental} and the {\it adjoint}. Fundamental
representations are used to describe the internal space
of fermions.  Adjoint
representations are used to describe the internal space
of bosons - the gauge vector fields. To each type of
representations singlets are added, in order to describe fermions and
bosons  which don't manifest
the colour or the weak charge.

Generators of the infinitesimal
transformations of the Lorentz group $ M^{mn} $ and the groups
defining charges $ \tau^{Ai} $

\begin{eqnarray}
M^{mn} & = & L^{mn} + \left\{ \begin{array}{l} {\cal S}^{mn}\\
S^{mn}\end{array} \right\},\; m,n \in \{0,1,2,3\},\\
\tau^{Ai} & = & \phantom{ L^{mn} +; } \left\{
\begin{array}{l} {\cal T}^{Ai}\\
T^{Ai} \end{array} \right\},\; A \in \{ 1,2,3 \}, i \in \{1,n_A\}.
\label{en2}
\end{eqnarray}

\noindent define representations, with $ n_A = 
N_A^{2} - 1 $ for $
SU(N_A)$ and  $ n_A = N_A^{2} $ for $U(N_A) $.

\noindent
For later convenience we introduce 
the generalized commutation relations \cite{man}

\begin{eqnarray}
 \{ A,B \} := AB - (-1) ^{n_{AB}} BA, 
\;\; n_{AB} = \left\{ \begin{array} {ll} +1, & {\rm if\; A\; and\; B \;
have\;Grassmann\; odd\; character}\\
0, &{\rm otherwise.} \end{array} \right\}
\label{en3}
\end{eqnarray}

\noindent  The infinitesimal generators of the
Lorentz group  $ SO(1,3) $:  $ M^{mn} $ fulfil the commutation 
relations\footnote[1]{
We use in this paper units in which $ c = 1 = \bar{h}$.}

\begin{eqnarray}
\{ M^{ab}, M^{cd} \} = i ( M^{ad} 
\eta^{bc} + M^{bc} 
\eta^{ad} - M^{ac} \eta^{bd} - M^{bd} \eta^{ac} ), 
\label{en4}
\end{eqnarray}

\noindent while $\tau^{Ai},$ which are the infinitesimal
generators of the groups $ 
SU(3) $, $SU(2) $ and $U(1)$, 
 fulfil the
commutation relations 

\begin{eqnarray}
\{\tau^{Ai}, \tau^{Bj}\} = if^{Aijk} \delta^{AB}\;\tau^{Ak}.
\label{en4}
\end{eqnarray}

\noindent All operators in Eqs.(4,5) have an even Grassmann
character. 

If we define\footnote[2]{ $\varepsilon_{a_1 a_2 ...a_n} =
\varepsilon^{a_1 a_2 ...a_n} $ is the 
totally antisymmetric 
tensor with $\varepsilon_{123..n} = 1$.} 
$ \;\;M_i^{\pm} = \frac{1}{2} (\frac{1}{2} \varepsilon_{ijk} M^{jk}
\pm i M^{0i}),\; i,j,k \in \{1,2,3 \}$, we find $ \{M_i^{\pm},$ $
M_j^{\pm}\} = i \varepsilon_{ijk} M_k^{\pm},\;\; \{ M_i^{\pm},$ $
M_j^{\mp} \} = 0$, which demonstrates the $ SU(2)\times SU(2) $
structure of the group $SO(1,3) $. It follows that $M^{ij} =
\varepsilon_{ijk} ( M^{k+} + M^{k-}), M^{0k} = -i( M^{k+} - M^{k-})$.
Operators $ M^{k+}$ and $ M^{k-}$ define the left handed and the
right handed representations  of the Lorentz group of either
fundamental or adjoint types.

Let ${\cal S}^{mn}$ and ${\cal T}^{Ai}$ stand for the operators,
defining the fundamental
representations of the Lorentz group and the group $ SU(N_A) $,
respectively, and let $ S^{mn}$ and
$ T^{Ai}$ stand for operators defining the adjoint
representations of the corresponding groups. 
Both types of operators fulfil the algebra of Eqs.(4,5),
respectively. 
Operators of the adjoint representations are determined by the
structure constants of the groups. To see this for the Lorentz
group $SO(1,3) $, we take into account the $
SU(2) \times SU(2) $ structure of  this group, presented above.
We find  $ (S^{ij})_{lm} = i 
\varepsilon^{ijk} \varepsilon_{klm}$, 
$ (S^{0i})_{lm} = i \varepsilon_{ilm} $. The operators, 
defining the adjoint representations of
the group $SU(N_A)$, are: $ (T^{Ai})_{jk} = -i f^{Aijk}$. 
The operators ${\cal
S}^{mn}$ ( as well as $ {\cal S}_{\pm}^k $ ) are  
the Pauli $2\times 2$ matrices, while $ S^{mn}$ ( as well as 
$  S_{\pm}^k $ )  are $3 \times 3$ matrices. Similarly, 
operators ${\cal T}^{Ai}$ are $N_A \times N_A$ matrices, while $
(T^{Ai})_{jk} = -i f^{Aijk}$ are $n_A \times n_A$ matrices.

The spin operators in the internal space  and the angular
momentum operators in the ordinary space time 
fulfil the same algebra ( $ M^{ab} $ from Eq.(4) is equal to
either ${\cal 
S}^{mn} $, or to $
S^{mn} $, or to $ L^{mn} $ ).  In order that the
corresponding group transformations 
are coupled, theories assume the same parameters for the
corresponding group elements.

{\it Modern theories} try to unify the internal spaces of
charges, but 
they {\it don't unify  the internal spaces of charges  with the
internal space of spins}, although it seems tempting to
generalize Eq.(1) in a d dimensional space time as follows:

$$ M^{ab} = L^{ab} + \left\{ \begin{array}{l} {\cal S}^{ab}\\
S^{ab}\end{array} \right\},\; a,b \in \{0,1,2,3,5,..,d\},
\eqno(1\grave{ }) $$ 
 
\noindent recognizing that for energies $\;\;\ll
\frac{1}{< x^h >}$, 
where $ < x^h >, \; h \in \{5,6,..,d \} $, is the radius of the
$ h-th$ 
coordinate, the contribution of $ L^{hk}$ to $ M^{hk}, \; h,k
\in \{ 5,6,...,d \} $ is nonnoticeable. Since we do not observe
more then four ordinary coordinates, d-4 coordinates should be
compactified.  At low enough energies then Eq.($1\grave{ }$)
manifests approximately as Eqs.(1,2), with

$$ \tau^{Ai} = {\it c}^{Ai}{ }_{hk} M^{hk} = {\it
c}^{Ai}{ }_{hk} \left\{ \begin{array}{l} {\cal S}^{hk}\\
S^{hk}\end{array} \right\},\; h,k \in \{5,..,d\},\; A
\in\{1,2,3\}, \;\; i = \in \{1,..,n_A \},  \;\;
{\it c}^{Ai}{ }_{hk} = - {\it c}^{Ai}{ }_{kh} \eqno(2 \grave{ }\;a) $$
 
\noindent and with coefficients $ {\it c}^{Ai}{ }_{hk} $ which 
fulfil the equation 

$$ -4 {\it c}^{Ai}{ }_{hk} {\it c}^{Bjk}{ }_{l} - \delta^{AB}
f^{Aijk} {\it c}^{Ak}{ }_{hl} = 0, \eqno(2 \grave{ }\;b) $$

\noindent so that for the operators $\tau^{Ai}$  the commutation
relations of  Eq.(5) are valid.

 This kind of unification of spins and charges was proposed in
ref.\cite{man,ana} within the approach that space time has d
ordinary commuting ( $ x^a x^b - x^b x^a = 0,\; a,b
\in\{0,1,2,3,5,..,d \} $) and 
d Grassmann anticommuting  ($ \theta^a \theta^b +
\theta^b \theta^a = 0,\; a,b \in \{ 0,1,2,3,5,..,d \}$)
coordinates \footnote[3]
{The metric tensor $ \eta
_{ab}$ $ = diag (1,$ $ -1, -1, -1,..., -1) $ lowers the indices of a
vector $\{ \theta^a \} = \{ \theta^0, \theta^1,..., \theta^d \},
\theta_a = \eta_{ab} \theta^b$. Linear transformation actions on
vectors
$ (\alpha \acute{\theta}^a + \beta \acute{x}^a ) = L^a{ }_b
(\alpha \theta^b + \beta x^b ),$
which leave forms
$ ( \alpha \theta^a + \beta x^a ) ( \alpha \theta ^b + \beta
x^b ) \eta_{ab} $
invariant, are called the Lorentz transformations.}, with $ d\ge
15$, and that consequently all 
symmetries of a system are connected with only coordinate
transformations in ordinary and Grassmann space.

It turns out \cite{man} that {\it in Grassmann space there exist
two kinds of operators of 
the Lorentz transformations: one defining spinorial kind of
representations,}  which include what is known as
fundamental representations, {\it the other defining vectorial
kind of representations,} which include what is known as adjoint
representations. {\it They are therefore appropriate to describe
spins and charges for fermions and bosons, respectively, unifying
spins with charges for each kind of representations separately.}  

In this paper we comment on fermionic and 
bosonic representations, defined by these two kinds of
operators, from the point of view of the Electroweak Standard
Model  and on  a "no go"
theorem\cite{col,haag}.  A more ellaborate version
will be published in a separate paper\cite{ana}.

\vspace{0.6cm}

Let us briefly present  the approach.

A linear vector space spanned over a Grassmann coordinate space of d
coordinates has the dimension $ 2^d$. If monomials $
\theta^{\alpha_1} \theta^{\alpha_2}....\theta^{\alpha_m} $
are taken as a set of basic vectors with $\alpha_j \neq
\alpha_k,$ half of the vectors have an odd (those with an odd m)
and half of the vectors  an even (those with an even m)
Grassmann character. Any vector in this space may be represented
as a linear superposition of monomials 
$ f(\theta) = \alpha_0 + \sum_{i=1}^{d}  \alpha _{a_1a_2 ..a_i}
\theta^{a_1} \theta^{a_2}....\theta^{a_i},\;\; a_k< a_{k+1},$ 
where constants $\alpha_0, \alpha_{a_1a_2..a_i}$ are complex
numbers. 

On this linear space we define the following linear operators
\cite{man}:

$$ p^{\theta} { }_a := i {\overrightarrow{{\partial}^{\theta}}}{
}_a , \;\;
 \tilde{a} ^a := i(p^{\theta a} - i \theta^a) ,\;\;
\tilde{\tilde{a}}{}^a := -(p^{\theta a} + i \theta^a). \eqno
(6) $$

\noindent According to Eqs.(3,6) we find

$$ \{p^{\theta a}, p^{\theta b} \} = 0 = \{ \theta^{a},
\theta^{b}\}, \{p^{\theta a}, \theta^{b}\} = -i \eta^{ab},
\{\tilde{a}^{a}, \tilde{a}^{b} \} = 2 \eta^{ab}
= \{\tilde{\tilde{a}}{ }^{a}, \tilde{\tilde{a}}{ }^{b} \}, \{
\tilde{a}^{a}, \tilde{\tilde{a}}{ }^{b} \} = 0.   \eqno(7)$$

\noindent ( We see that $\theta ^a $ and $ p^{\theta a} $ form a
Grassmann 
odd Heisenberg algebra, while $ \tilde a^a $ and $
\tilde{\tilde{a}}{ }^a $ form the Clifford algebra.)

Any of the two   bilinear
forms\cite{man} 

$$ \tilde{\cal S}^{ab}: = - \frac{i}{4} [\tilde{a}^a , \tilde{a}^b
], \;\; \tilde {\tilde{\cal S}} { }^{ab}: = - \frac{i}{4} [ \tilde
{\tilde a}{ }^a , \tilde {\tilde a}{ }^b ] , \eqno(8a)$$ 

\noindent with $ [A, B]:= AB - BA,$ close the algebra of the
Lorentz group 
$ SO(1, d-1 ) $ ( Eq.(4)) and define what we call {\i the
spinorial representations} of 
the Lorentz group and of subgroups of the Lorentz
group.\cite{man,ana} We  use these representations to
describe spins and 
charges of fermionic fields. In this paper we shall make use of only 
$\tilde{\cal S}^{ab} $. We shall therefore omit the sign
($\tilde{ } $) and write $a^a$ and
${\cal S}^{ab}$ for the corresponding operators, using the
symbol, which we introduced for operators, decribing the
fundamental representations.

The operators

$$ S^{ab} : = ( \theta^a p^{\theta
b} - \theta ^b p^{\theta a} )                     \eqno  (8b)$$

\noindent close the algebra of Eq.(4) and define  
what we call {\it the vectorial }
representations of the Lorentz group $ SO(1,d-1)$
and of subgroups of this group.\cite{ana} We use these
representations to describe spins
and charges of bosonic fields. For the vectorial operators the
symbol is used, which was introdused for operators of the
adjoint representations.

Both kinds of operators are, according to Eqs.(6), bilinear
forms of differential operators.



It can be proved for  $d=2n$, where $n$ is an integer,  that
$ M^2,\;\;M^2 = \frac{1}{2} M^{ab} M_{ab}, $ and $\Gamma,\;\;
\Gamma = \frac{i(-2i)^{n} }{(2n)!} $ $
\varepsilon_{a_1a_2...a_{2n}}$ $ M^{a_1a_2} ....M^{a_{2n-1}a_{2n}},$
are among invariants of the Lorentz group: 
$\;\; \{ M^2, M^{cd} \} = 0,\;\; 
\{ \Gamma, M^{cd} \} = 0.\;\;$

According to Eqs.($1\grave{ }, 2 \grave{ }a, 2 \grave{ }
b $) the algebra of the group $ SO(1,14) $ 
contains as subalgebras the algebras of subgroups $
SO(1,4) $ and $ SO(10)$. The group $ SO(1,4) $ rather then 
the group $SO(1,3)$ will be used to
describe the spin of fermionic and bosonic fields, the group $
SO(10) $, containing 
subgroups $ SU(3), SU(2), U(1), $ will be used to describe
the charges of fermionic and bosonic fields.
The generators of  $SO(1,3)$, which is the subgroup of
 $SO(1,4)$, determine  spins of  fermionic ($
{\cal S}^{mn}, \;\;m,n \in \{0,1,2,3 \},$) and of bosonic (
$ S^{mn},\;\;m,n \in \{0,1,2,3 \}, $) fields.
The remaining generators of group 
$ \; SO(1,4), \;\; $ that is $ M^{5m} , \; m \in \{ 0,..,3 \}
\; $, will be denoted by a special
name $ \gamma^a: = -2i M^{5a} $. 
In the case of generators of spinorial
character, $ {\gamma}^m = -2i  S^{5m} = a^5 a^m $ may be
recognized as the Dirac $\gamma^m $ matrices, 
with all the desired properties
\footnote[4]{ Operators $ a^a $ are
Grassmann odd operators. Operating on spinors they  change
fermions to bosons, changing the Grassmann character from odd to
even, and therefore $ a^m$ can not be recognized as Dirac
$\gamma ^m $ matrices \cite{man}. One also finds $ 
\Gamma^{(4)} = i a^0 a^1 a^2 a^3 = i \gamma^0  \gamma^1  \gamma^2 \gamma^3
$, where exponent (4) of $\Gamma^{(4)}$ denotes the four
dimensional subspace.}.

Looking for the $ SU(3) \times SU(2) \times
U(1) $ structure of the group $ 
SO(10) $ in accordance with Eqs.($ 2 \grave{ }a, 2 \grave{
 }b $), where
operators $\tau^{1i}, \tau^{2i}, \tau^{3i} $ close the
subalgebras according to Eq.(5)
and the coefficients $ f^{1ijk} $, $\varepsilon^{ijk} $ are the
structure constants of the groups $ SU(3) $, $ SU(2) $,
respectively, one finds:

$$ \tau^{1\;1} := \frac{1}{2}\; (M_{6\; 9} - M_{7\; 8}),\;\;\;\;
   \tau^{1\;2} := \frac{1}{2}\; (M_{6\; 8} + M_{7\; 9}),\;\;\;\;
   \tau^{1\;3} := \frac{1}{2}\; (M_{6\; 7} - M_{8\; 9}),$$
$$ \tau^{1\;4} := \frac{1}{2}\; (M_{6\; 11} - M_{7\; 10}),\;\;\;\;
   \tau^{1\;5} := \frac{1}{2}\; (M_{6\; 10} + M_{7\; 11}),\;\;\;\;
   \tau^{1\;6} := \frac{1}{2}\; (M_{8\; 11} - M_{9\; 10}),$$
$$ \tau^{1\;7} := \frac{1}{2}\; (M_{8\; 10} + M_{9\; 11}),\;\;\;\; 
\tau^{1\;8} := \frac{1}{2{\sqrt 3}}\; (M_{6\; 7} + M_{8\; 9} - 2 M_{10\;
11}), $$

$$ \tau^{2\;1}: =  \frac{1}{2} \;(M_{12\; 15} - M_{13\; 14}),\;\;\;\;
   \tau^{2\;2}: =  \frac{1}{2} \;(M_{12\; 14} + M_{13\; 15}),\;\;\;\;
\tau^{2\;3}: = \frac{1}{2} \;(M_{12\; 13} - M_{14\; 15}),
\eqno(9)$$

$$ \tau^{3\;1}: = \sqrt{\frac{3}{5}}\;\; [- \frac{1}{3}\; (M_{6\; 7} +
M_{8\; 9} + M_{10\; 11}) + \frac{1}{2} \;(M_{12\; 13} + M_{14\; 15})]. $$

\noindent Operators $ \tau^{A i},\; A \in\{1,..,3\}, \; i
\in\{1,n_A\}, $ define either 
spinorial ($ M^{hk} = {\cal S}^{hk}, \;\;\tau^{A i} = {\cal
T}^{A i} $) or vectorial 
(  $ M^{hk} = S^{hk}, \;\;\tau^{A i} = T^{A i} $)
representations.

To find the irreducible representations of the  group $
SO(1,14)$ in terms of subgroups $ SO(1,4) \times SU(3) \times SU(2)
\times U(1) $, the eigenvalue problem for the Casimir
operators and all the commuting
operators for each of  subgroups has to be solved:

$$ <\theta|{\cal A}_i|\varphi> = a^f{ }_i
<\theta|{\varphi}> ,\;\;
<\theta|A_i|\phi> = a^b{ }_i <\theta|\phi>,\;\;i = \{1,r\}
,\eqno(10)$$ 

where $ {\cal A}_i $ and $ A_i $ stand for $r$ commuting
operators 
of spinorial and vectorial character, respectively and $
a^f{ }_i $ and $ a^b{ }_i$ for the coresponding eigenvalues.

To solve  Eqs.(10), one has to express the operators in the
coordinate representation  and write the
eigenvectors 
as polynomials \footnote[6]{To orthonormalize vectors, the inner
product has to be 
defined\cite{man}.}$\;$  of $\theta ^a,\;\; a = 1,15 $. 
We assume that spinorial representations have an odd and
vectorial an even Grassmann character\footnote[7]{ $\;$In the
cannonical quantization of fields spinorial 
representations should quantize to fermions, vectorial to
bosons.},$\;\;$ respectively.

According to Eqs.($1\grave{ }, 2 \grave{ }a, 2 \grave{ }b$), one
can first solve the eigenvalue 
problem separately in each of subspaces in which generators of the
groups $SO(1,4)$, $SU(3)$ and $SU(2)$ operate, respectively, and
then get the
representations in the hole  space as the direct
product of representations in different subspaces.

In the spinorial case one finds\cite{ana}
eight bispinors, four left and four right handed.
The operators $\gamma^m$ connect them into four four spinors. Each
can be a triplet or a singlet with respect to $SU(3)$, a doublet
or a singlet with respect to $SU(2)$, while the possible values for
the $U(1)$ charge are presented on TableI.

\vspace{3mm}

\begin{center}
\begin{tabular}{|c|c|c|}
\hline
$<\theta|{\varphi}^{a}_{i}>$&
$<\theta|{\varphi}^{b}_{j}>$& 
${\cal T}^{31} \times \sqrt{ \frac{5}{3}}$\\
\hline 
$ triplets$ & $ doublets $ & $\pm 
\frac{1}{6}$\\
\hline 
$ triplets$ & $ singlets $ & $\pm 
\frac{2}{3}, \pm \frac{1}{3}$\\ 
\hline
$ singlets$ & $ doublets $ & $ \pm 
\frac{1}{2}$\\
\hline 
$ singlets$ & $ singlets $ & $0, \pm 1 $\\ 
\hline
\end{tabular}

\end{center}

\noindent Table I. The eigenvalues of the spinorial operator
${\cal 
T}^{31}$ forming the algebra of
the group of $U(1)$ (Eq.(9)), with $M^{ab} = {\cal
S}^{ab}, 
$ for the representations, which are the direct products of the
spinorial representations of the group $SU(3)$ and the group
$SU(2)$. Index $a$ runs over different triplets or singlets
belonging to the group $SU(3)$ and
index $i$ runs within the same triplet. Index $b$ runs over
different doublets or singlets of the group $SU(2)$ and $j$ within
the same doublet\cite{ana}.

\vspace{0.1cm}

For   the vectorial case, one finds\cite{ana}
a scalar, a pseudoscalar and two three vectors, one left and one
right handed, and two four vectors. Each of them can be an
octet, a triplet or a singlet with respect to $SU(3)$, or a
triplet, a doublet or a singlet with respect to $SU(2)$, while the
possible values for the $U(1)$ charge are presented on Table II.

\vspace{1mm}

{\it For $d = 15$, Grassmann space offers all the representations
needed in the Electroweak Standard Model to describe fermions,
gauge fields and
Higgs scalars\cite{ana}. }

We find left handed spinors, which are $SU(3)$
triplets and $SU(2)$ doublets with $U(1)$ charge equal to $\pm
\frac{1}{6}$  and right handed spinors, which
are $SU(3)$ triplets and $SU(2)$ singlets with $U(1)$ charge
equal to $\pm \frac{2}{3}$ and $\mp \frac{1}{3}$, needed to
describe quarks, left handed spinors, which are $SU(3)$
singlets and $SU(2)$ doublets with $U(1)$ charge equal to $\mp
\frac{1}{2}$ and right handed spinors, which are $SU(3)$ singlets
and $SU(2)$ singlets with $U(1)$ charge equal to $\mp 1$, needed
to describe leptons. Since there are four four spinors, the
approach predicts, if quarks and leptons are elementary fields,
four rather then three families.

\begin{center}
\begin{tabular}{|c|c|l|}
\hline
$<\theta|{\phi}^{a}_{i}>$& $<\theta|{\phi}^{b}_{j}>$&
${T}^{31} \times \sqrt{ \frac{5}{3}}$\\
\hline
$ octets$ & $ triplets $ & $0$\\
\hline
$ octets$ & $ doublets $ & $\pm \frac{1}{2}$\\
\hline
$ octets$ & $ singlets $ & $0,\pm 1$\\
\hline
$ triplets$ & $ triplets $ & $\pm \frac{1}{3}, \pm \frac{2}{3}$\\
\hline
$ triplets$ & $ doublets $ & $\pm \frac{1}{6}, \pm \frac{5}{6}$\\
\hline
$ triplets$ & $ singlets $ & $\pm \frac{1}{3}, \pm \frac{1}{2},
\pm \frac{2}{3}, \pm \frac{3}{2}, \pm \frac{5}{3}$\\
\hline
$ singlets$ & $ triplets $ & $0$\\
\hline
$ singlets$ & $ doublets $ & $ \pm \frac{1}{2}$\\
\hline
$ singlets$ & $ singlets $ & $0,\pm 1$\\
\hline
\end{tabular}
\end{center}

\noindent Table II. Eigenvalues of the vectorial operator
$T^{31}$ from Eqs.(9),
with $M^{ab} = S^{ab}, $ for the vectorial
representations which are the direct product of representations
of the group $SU(3)$ and the group $SU(2)$. Index $a$ runs
over different octets, triplets and singlets, index $i$ within the
octet and the triplet. Index $b$ runs over different triplets,
doublets and singlets, $j$  within the triplet or the doublet.

\vspace{0.3cm}

We find left  and right handed three vectors, which are $SU(3)$
octets and $SU(2)$ singlets with $U(1)$ charge equal zero,
needed to describe gluons, left  handed three vectors,
which are $SU(3)$ singlets and $SU(2)$ triplets with $U(1)$
charge equal zero, needed to describe weak bosons and left and
right handed three vectors, $SU(3)$
 and $SU(2)$ singlets with 
$U(1)$ charge equal zero, needed to
describe the $U(1)$ field.  

We find also scalars,
which are $SU(3)$ singlets and $SU(2)$ doublets, needed to
describe Higgs fields.

In this approach there is more representations than needed in the
Electroweak Standard Model.  The detailed study of those will be
presented elsewhere\cite{ana}. 
The structure of the Grassmann space, with the limited number of
vectors, however, limits the possible representations allowed by the
group theory, offering only representations of groups, which are
subgroups of the group $SO(1,d-1)$.     

Let us point out that among the representations of $SO(1,14)
\supset SO(1,3) \times SU(3) \times SU(2) \times U(1) $ one can
find no bosinos, which would be 
$SU(3)$ octets or $SU(2)$ triplets, required by 
supersymmetric extensions of the Electroweak Standard Model. These
models assume the existence of 
fermions, which are in the adjoint representations with respect
to  groups determining charges. They  require also the existence
of sfermions, that is bosons, which are in the 
fundamental representations with respect to  groups
determining charges. 

It looks like that {\it unification of spins and 
charges doesn't support} the {\it simplified version of a
supersymmetry}, unless bosinos are constituent fields, which would
mean that all known fields are  constituent
fields as well. 

The proposed 
approach does manifest supersymmetry\cite{man}: there are equal
number of Grassmann odd ($2^{d-1}$) and Grassmann even
representations($2^{d-1}$).

Gravitons 
and gravitinos  appear as tensor fields\cite{man}.

The approach, in which spins and charges unify, suggests the
unification of all interactions: Yang-Mills fields with gravity.
To see this, let us look at the equation of motion for a
massless spinorial particle in  d dimensional ordinary and d
dimensional Grassmann space in the presence of a gravitational
field\cite{man}: 

$$ \gamma^a p_{0a} = 0,\;\;\; p_0{ }^a p_{0a} = 0, \eqno(11) $$ 

\noindent with

$$ p_{0a} = f^\mu{ }_a p_{0 \mu},\;\; p_{0 \mu} = p_{ \mu} -
\frac{1}{2} {\cal S}^{a b} \omega_{a b \mu} \;,\; 
\omega_{a b \mu}=\frac{1}{2} (e_{a \mu b} - e_{b \mu a}).
\eqno(11a) $$

\noindent Vielbeins $ e^a{ }_{\mu}$ and their inverses $
f^{\mu}{ }_a, \;\; 
e^{a}{ }_{\mu} f^{\mu}{ }_b = \delta^a { }_b \;,\; 
 f^{\mu}{ }_{a}  e^{a} { }_{\nu} = \delta^{\mu} {
}_{\nu}$, and spin connections $ e^a{ }_{\mu b} $ depend on
ordinary on on Grassmann coordinates.
The detailed derivation of the above equation is presented in
refs.\cite{man}. 

Under {\it special conditions}, when vielbeins have a block
structure

$$ \left( \begin{array}{cc|cc}
   e^{m}{ }_{\alpha}& & & 0\\
   & & &\\ \hline
   & & &\\
   0 & & & e^{h}{ }_{\sigma}
\end{array} \right), \;\;\;\;
\alpha,m \in (0,..,3),\: \sigma,h \in (5,..,d), \eqno
(11b)$$ 

\noindent and depend only on $x^{\alpha}$ and $\theta^a$,
$\alpha \in \{0,..,3\},$  while spin connetions $\omega_{ab
\alpha} $ fulfil the equations $\omega_{hk \alpha} = 2 {\it
c}^{Ai}{ }_{hk} A^{Ai}_\alpha,\;\; h,k \in
\{6,..,d \},\;\; \alpha \in\{0,..,3\} $, which according to
Eqs.($2\grave{ }$) means that most of 
$\omega_{hk \alpha}$ are equal to zero, while those nonzero
values are expressible with gauge fields ${\cal A}^{Ai}{
}_{\alpha}$ (the number of these fields is
$12$ for each $\alpha$: $8$ for $SU(3)$, $3$ for $SU(2)$ and $1$
for $U(1)$), the gravitational field in d dimensional ordinary
and d dimensional
Grassmann space time manifests in the four dimensional
(sub)space time
as an ordinary gravity and Yang-Mills gauge fields\footnote[8]{
$\;$ Let us point out that the vielbein structure from Eq.(11b)
is not the one proposed by ordinary Kaluza-Klein
theories\cite{wit}. In Kaluza-Klein 
theories the 
nondiagonal vielbeins 
determine Yang-Mills fields.

}. 

We find\cite{man}

$$  \gamma^a f^{\mu}{ }_a p_{0 \mu} =  \gamma
^m f^{\alpha}{ }_m ( p_{ \alpha} - \frac{1}{2} {\cal S} ^{mn}
 \omega_{mn \alpha} +{\cal A}_{\alpha} ), \;\;
{ \rm where}\;\; {\cal A}_{\alpha} = \sum_{A,i} { \cal T}^{Ai} {\cal
A}^{Ai}_{\alpha}, 
\eqno(11c) $$ 

\noindent with
$\sum_{A,i} { \cal T}^{Ai} {\cal A}^{Ai}_{ \alpha} = 
\frac{1}{2}  {\cal S}^{hk}  \omega_{hk \alpha} ,
 \;\;h,k = 6,..,d. $

For $e^{m}{ }_{\alpha} = \delta ^m{ }_{\alpha}$ one easily sees
that Eq.(11) manifests 
the Dirac equation for a particle, whose spin is determined by
${\cal S}^{mn},\; m,n\in \{0,..,3\} $ and whose Yang-Mills
charges are determined by $ {\cal S}^{hk},$ in the
presence of only gauge  fields.

In the proposed theory charges as well as spins are
determined by the generators of the Lorentz transformations in
Grassmann space.  
Spin connections (Eq.(11c)) rather than vielbeins determine
gauge fields.

Since this paper suggests the unification of spins and charges
within the group $SO(1,14)$,  and since the paper of Colemann and
Mandula\cite{col} together with the paper of Haag,
$\bar{L}$opusza\' nski and Sohnius\cite{haag} speaks against it,
convincing the physical
community that it is no hope for this kind of unification, let
us comment on this "no go" theorem. 

To assure the reader
that there is no contradiction between the proposed unification
and the "no go" theorem it is only needed to say that the
dynamics of fields in our approach is defined in a d dimensional
ordinary ( 
and d dimensional Grassmann) space so that the scattering
matrix, defined in a similar way as in Ref.\cite{col}, 
is unitary in d rather then in four ordinary dimensions, as
assumed in the "no go" theorem\cite{col,haag}. 
If all the coordinates but  four  
are compactified, as we  already have assumed, then at energies (of
scattering particles in a center of mass coordinate system),  low
compared to the inverse radii of the subspace of d-4 dimensions,
the S - matrix
manifests approximately as an unitary matrix in a four
dimensional subspace, and as an analytic function
of only the four momenta $p^m,\; m\in \{0,1,2,3 \}$,  with the
connected symmetry group isomorfic to the direct product of the
Poincar\' e group in four dimensions and the groups defining
charges in d-4 Grassmann dimensions
\footnote[8]{$\;$
Situation is similar to the
nonrelativistic approach in the four 
dimensional space-time:
Instead of the Poincar\' e group, the group of rotations
$SU(2)$, the group of
translations in the three dimensional subspace, the internal
group of charges  and the Galilean symmetry  are  assumed to define
the exact symmetry of the S matrix, which is assumed to be unitary
and an analytic function of the center of mass energy of a two
particle state and in which time plays a role of parameter
rather then a coordinate. The manifestation of the generators
$M^{0i},\; i \in \{ 1,2,3 \},$ 
is supposed to be negligible.
}. 
At such
energies the Poincar\' e group in a d dimensional space
manifests
as the direct product of the
Poincar\' e group in a four dimensional subspace and the groups
describing charges in the way we have commented
above.
This is what the theorem
of Colemann and Mandula states for: If the
Poincar\' e transformations in four dimensional space should not
transform one charge degree of freedom into another ( Coleman
and Mandula speak about particle types or different irreducible
representations of the Poincar\' e group ), the generators of
the Poincar\' e group and the internal group should commute.

With the 
growing energy, however, not only the S matrix would start to
manifest the unitarity in the d dimensional space, but 
charges and spins start to manifest as a part of the
Lorentz group  in the d dimensional ordinary and
Grassmann space.



\section{Acknowledgement. } This work was supported by Ministry of 
Science and Technology of Slovenia. 
The author appreciates fruitful discussions with H.B. Nielsen.

\end{document}